# First-principles Prediction of Potential Candidate Materials $M$Cu$_3$X$_4$ ($M$ = V, Nb, Ta; $X$ = S, Se, Te) for Neuromorphic Computing

Baoxing Zhai,[1] Ruiqing Cheng,[1] Tianxing Wang,[2] Li Liu,[1] Lei Yin,[1] Yao Wen,[1] Hao Wang,[1] Sheng Chang,[1] and Jun He[1,3,*]

[1] *Key Laboratory of Artificial Micro- and Nano-structures of Ministry of Education, and School of Physics and Technology, Wuhan University, Wuhan 430072, China*
[2] *School of physics, Henan Normal University, Xinxiang 453007, China*
[3] *Wuhan Institute of Quantum Technology, Wuhan 430206, China*

Inspired by the neuro-synaptic frameworks in the human brain, neuromorphic computing is expected to overcome the bottleneck of traditional von-Neumann architecture and be used in artificial intelligence. Here, we predict a class of potential candidate materials, $M$Cu$_3$X$_4$ ($M$ = V, Nb, Ta; $X$ = S, Se, Te), for neuromorphic computing applications through first-principles calculations based on density functional theory. We find that when $M$Cu$_3$X$_4$ are inserted with Li atom, the systems would transform from semiconductors to metals due to the considerable electron filling [~0.8 electrons per formula unit (f.u.)] and still maintain well structural stability. Meanwhile, the inserted Li atom also has a low diffusion barrier (~0.6 eV/f.u.), which ensures the feasibility to control the insertion/extraction of Li by gate voltage. These results establish that the system can achieve the reversible switching between two stable memory states, i.e., high/low resistance state, indicating that it could potentially be used to design synaptic transistor to enable neuromorphic computing. Our work provides inspiration for advancing the search of candidate materials related to neuromorphic computing from the perspective of theoretical calculations.

---

[*]he-jun@whu.edu.cn

## I. INTRODUCTION

Industrial 4.0, also known as the era of intelligence, was put forward in 2014 [1], and artificial intelligence is expected to play an essential role in it. In recent years, many novel concepts have emerged around artificial intelligence, including neuromorphic computing [2–5], machine learning [6,7], and deep learning [8,9], etc. The research in these fields covers innovations in materials, electronic device structures and program algorithms. For neuromorphic computing, it is considered to be an advanced brain-like computing method with the advantages of energy saving and high efficiency. At present, a variety of schemes have been proposed to realize neuromorphic computing, such as memristors [2,4], ion intercalation [10–13], structural phase transition [14–18], heterostructure engineering [19,20], and spintronic devices [21–24]. Among them, ionic synaptic transistors can be designed based on ion intercalation. As shown in Fig. 1(a) and 1(b), the structure of ionic synaptic transistors is basically the same as that of the traditional metal-oxide-semiconductor field-effect transistors (MOSFETs). The biggest difference is that ionic synaptic transistors replace the oxide insulating layer in MOSFETs with an electrolyte layer, allowing them to work differently. As shown in Fig. 1(c), when a positive gate voltage is applied, metal ions in the electrolyte layer of ionic synaptic transistors will enter the channel layer, causing a change in channel conductance. While in a MOSFET, the gate voltage is applied to regulate the movement of carriers. Taking the n-type semiconductor substrate as an example [Fig. 1(d)], when a positive gate voltage is applied, an electron accumulation layer will be formed in the channel. Moreover, after the gate voltage is removed, the MOSFET will immediately return to its original state, while for the ionic synaptic transistor, because the intercalated particles are retained in the channel layer, the regulated conductance can be maintained, i.e. nonvolatility. Zhu *et al.* [12] have experimentally demonstrated that the ionic synaptic transistors based on two-dimensional (2D) van der Waals materials possess excellent application prospect in neuromorphic computing. Their work shows that as metal ions have two diffusion modes in 2D layered materials, one is adsorbed on the surface of the material and the other is embedded in the material, the devices can generate both short-term and long-term signals [12]. However, the diffusion mode that across the layered materials themselves may induce structural distortions or defects, which may challenge structural stability. Additionally, it has been reported that the structural transition of $2H$ phase and $1T'$ phase of $MoS_2$ can be realized by controlling the concentration of intercalated $Li^+$ ions [10].

These results undoubtedly add the uncertainty that synaptic transistors based on 2D layered materials may face in neuromorphic computing.

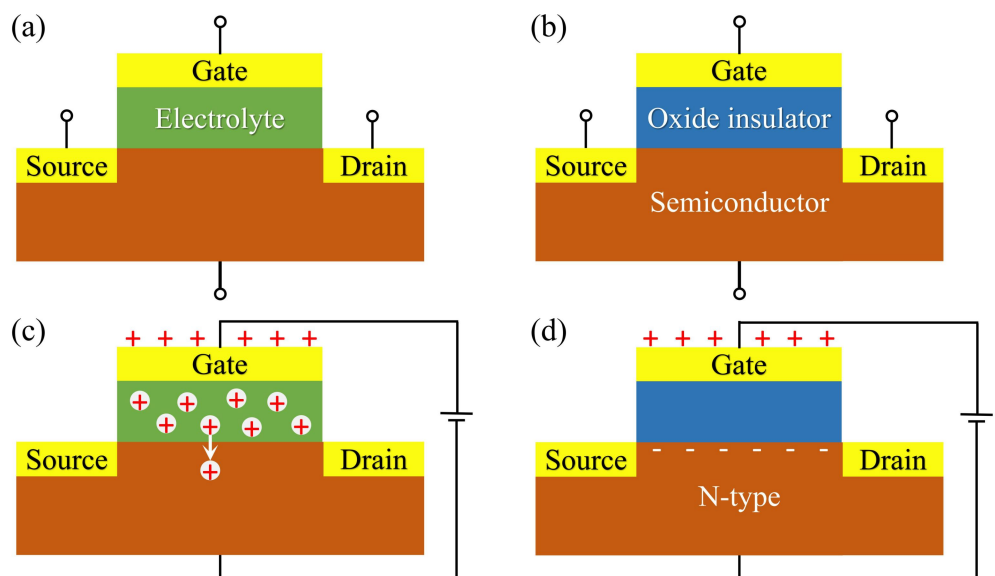

FIG. 1. Schematic diagram of the structures of (a) ionic synaptic transistors and (b) metal-oxide-semiconductor field-effect transistors (MOSFETs). Schematic diagram of the working principles of (c) ionic synaptic transistors and (d) MOSFETs.

Nowadays, with the rapid improvement of computer computing power, people can quickly screen out abundant satisfactory materials with specific properties from the database by setting some indicators and conducting high-throughput search [25–27], which can greatly accelerate the development and application of new materials. Similarly, as the hardware carrier of neuromorphic computing, it is also vital to find and screen suitable materials for the development of neuromorphic computing. However, due to the diversity and complexity of current schemes for neuromorphic computing, it is still quite tough to screen and predict materials by high-throughput computing or machine learning methods. In fact, to date, most materials studies related to neuromorphic computing are started from experiments, which undoubtedly limits the rapid development of the field.

Here, we note that $M$Cu$_3$X$_4$ ($M$ = V, Nb, Ta; $X$ = S, Se, Te), a class of semiconductor materials with a cubic lattice, have $P\bar{4}3m$ symmetry [Fig. 2(a) and 2(b)]. Due to the suitable range of band gaps, these kind of materials possess a great photoelectric application prospect [28–31]. In addition, they are theoretically predicted to have favorable thermoelectric properties [32,33]. We find that the structures of $M$Cu$_3$X$_4$ exist an intriguing feature, which is that there are no atoms in the body center and face center of their unit cells, just like an open framework. By contrast, fullerenes [34], a well-known class of carbon materials, are like a closed box. Numerous studies have shown that filling fullerenes with atoms may enable them to acquire some novel physical properties [35–37]. As early as 1978, Arribart *et al.* [38] have studied VCu$_3$S$_4$ embedded copper ions and pointed out that its crystal structure has the potential for intercalation research. Inspired by these studies and the above studies on neuromorphic computing, we speculate that $M$Cu$_3$X$_4$ may be applied to neuromorphic computing by inserting atoms. However, until now, no one had done the research.

In this work, we explore the potential of $M$Cu$_3$X$_4$ for neuromorphic computing through first-principles calculations. First, via structure optimization and transition state search, we ascertain the crystal structure of the ground state with different inserted atoms (H/Li/Na/K) and the diffusion barrier of the inserted atoms. Then, the structural stability after inserting different atoms is compared by phonon spectra. By the above analysis, we confirm that Li atom is more suitable for inserting into $M$Cu$_3$$X$$_4$. Subsequently, band structures calculations and charge transfer analysis show that the insertion of Li atom can transform $M$Cu$_3$$X$$_4$ from semiconductors to metals due to the atomic insertion-induced electron filling. Finally, we quantitatively analyze the resistance ratio of VCu$_3$S$_4$ and Li-VCu$_3$S$_4$ via electrical transport simulation, which can reach a surprising magnitude of $10^{12}$. These results signify that $M$Cu$_3$$X$$_4$ meet the requirements of ionic synaptic transistors reported in Ref. [11], thus we believe that they have potential applications in the device related to neuromorphic computing.

## II. COMPUTATIONAL METHODS

We performed first-principles calculations implemented in the Vienna Ab-initio Simulation Package (VASP) based on density functional theory (DFT) [39]. The exchange correlation potential is described with the Perdew-Burke-Ernzerhof (PBE) of the generalized gradient approximation (GGA) [40]. The PBE+U based on the approach of Dudarev *et al.* [41] was adopted for Cu 3$d$ electrons with U = 5.2 eV [28]. The electron-ion potential is described by the projected augmented wave (PAW) [42]. The kinetic energy cutoff of the plane wave was set to be 500 eV for the plane wave expansion. The Brillouin zone integration was carried out using 5×5×5 Monkhorst-Pack *k*-point meshes for geometry optimization of *M*Cu$_3$*X*$_4$ [43]. All geometric structures were fully relaxed until energy and forces are converged to 10$^{-6}$ eV and 0.01 eV/Å, respectively.

We studied the transport properties of *M*Cu$_3$*X*$_4$-based ionic synaptic transistors using Atomistix Toolkit-Virtual Nanolab (ATK-VNL) 2017 package with DFT and nonequilibrium Green function methods [44,45]. The electron exchange correlation is treated by GGA-PBE same as the calculation in VASP. The structural relaxations of each devices system are performed in advance and allowed until the absolute value of force acting on each atom is less than 0.05 eV/Å. The mesh cutoff was set as 75 Hartree, and the k-points grid 11×11×101 is used to sample the Brillouin zone of the electrodes in the *x*, *y* and *z*(transport direction) directions, respectively. The temperature of electrodes was set to 300 K to accelerate the transport calculation.

The non-equilibrium electron transport currents of state1 [Fig. 8(b)] at finite bias voltages were calculated by the Landauer–Buttiker formula:

$$I = \frac{e}{h} \int_{-\infty}^{\infty} \{T(E, V_b)[f_L(E - \mu_L) - f_R(E - \mu_R)]\} \, dE,$$

where $T(E, V_b)$ is the transmission coefficient under a finite bias $V_b$, $f_{L/R}(E - \mu_{L/R})$ corresponding to the Fermi-Dirac distribution for the left/right electrode, and $\mu_{L/R} = E_F \pm \frac{1}{2}eV_b$ represents the electrochemical potential of the left/right electrode. The I-V curve of state 2 [Fig. 8(b)] was calculated by integrating the zero-bias transmission spectrum in an increasingly wide bias interval.

### III. RESULTS AND DISCUSSIONS

The calculated geometric parameters (Table I) and band structures (Fig. S1 in the Supplemental Material (SM) [46]) of $M$Cu$_3$X$_4$ are in accordance with previous report [28]. We herein try four kinds of atoms H/Li/Na/K, and select the center of the unit cell as an insertion site, as shown in Fig. 2(c). The results show that inserting atom at the central site does not alter the lattice type and symmetry of $M$Cu$_3$X$_4$, except for the lattice constants, see Table I. Fig. 2(d) plots the lattice expansion rates of $M$Cu$_3$X$_4$ after inserting different atoms, which is obtained by the formula $\eta = \frac{a_1 - a_0}{a_0}$, where $a_0$ are the original lattice constants of $M$Cu$_3$X$_4$, and $a_1$ are the lattice constants after inserting atom. It can be seen that the lattice expansion rate will gradually rise with the atomic radius of the inserted atom.

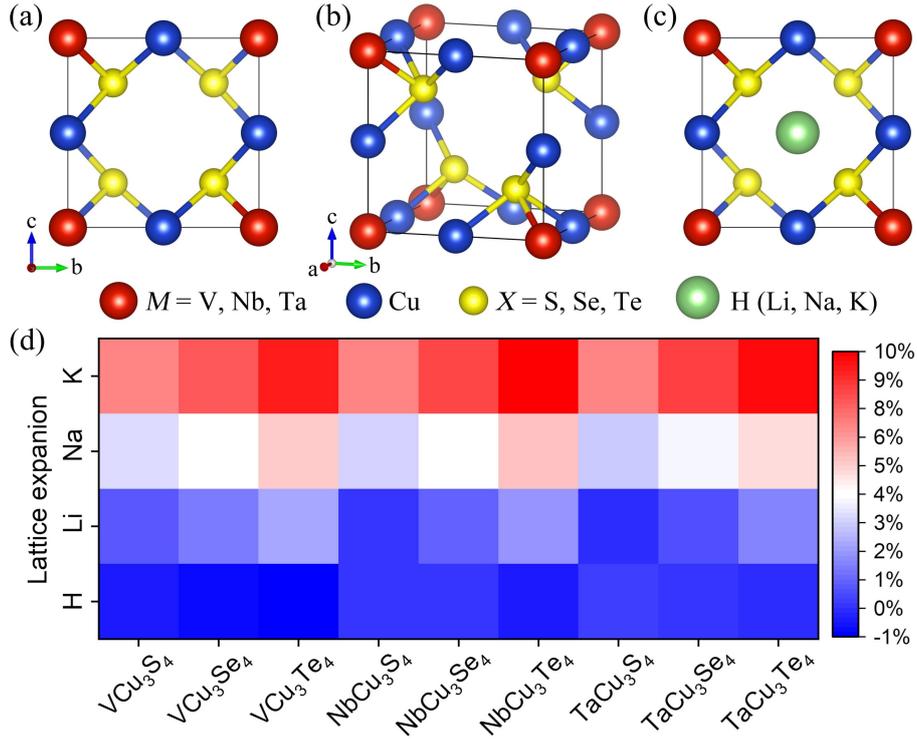

FIG. 2. (a) Side view of the $M$Cu$_3$X$_4$ unit cell alone the $a$ axis, same as that along the $b$ and $c$ axes. (b) 3D structural view of the the $M$Cu$_3$X$_4$ unit cell. (c) Schematic diagram of inserting the atom at the central site of the $M$Cu$_3$X$_4$ unit cell. (d) Calculated lattice expanion rates induced by inserting different atoms in $M$Cu$_3$X$_4$.

TABLE I. Calculated lattice constants for isolated $M$Cu$_3$X$_4$ ($a$) and inserting H ($a_H$), Li ($a_{Li}$), Na ($a_{Na}$), and K ($a_K$) atoms into $M$Cu$_3$X$_4$, and band gap $E_g$ of $M$Cu$_3$X$_4$ using PBE method. The lattice constants in parentheses are obtained by PBE+U method.

| Compound | $a$ (Å) | $a_H$ (Å) | $a_{Li}$ (Å) | $a_{Na}$ (Å) | $a_K$ (Å) | $E_g$ (eV) |
|---|---|---|---|---|---|---|
| VCu$_3$S$_4$ | 5.437 (5.464) | 5.421 | 5.487 (5.512) | 5.631 | 5.815 | 1.141 |
| VCu$_3$Se$_4$ | 5.649 (5.675) | 5.609 | 5.742 (5.762) | 5.896 | 6.092 | 0.887 |
| VCu$_3$Te$_4$ | 5.954 (5.983) | 5.905 | 6.108 (6.129) | 6.285 | 6.490 | 0.579 |
| NbCu$_3$S$_4$ | 5.540 (5.579) | 5.550 | 5.557 (5.572) | 5.726 | 5.925 | 1.905 |
| NbCu$_3$Se$_4$ | 5.726 (5.749) | 5.735 | 5.791 (5.816) | 5.975 | 6.190 | 1.505 |
| NbCu$_3$Te$_4$ | 6.013 (6.049) | 6.001 | 6.147 (6.171) | 6.351 | 6.592 | 1.030 |
| TaCu$_3$S$_4$ | 5.552 (5.586) | 5.574 | 5.555 (5.575) | 5.731 | 5.938 | 2.194 |
| TaCu$_3$Se$_4$ | 5.725 (5.791) | 5.736 | 5.771 (5.804) | 5.963 | 6.198 | 1.820 |
| TaCu$_3$Te$_4$ | 6.020 (6.054) | 6.020 | 6.133 (6.167) | 6.331 | 6.585 | 1.242 |

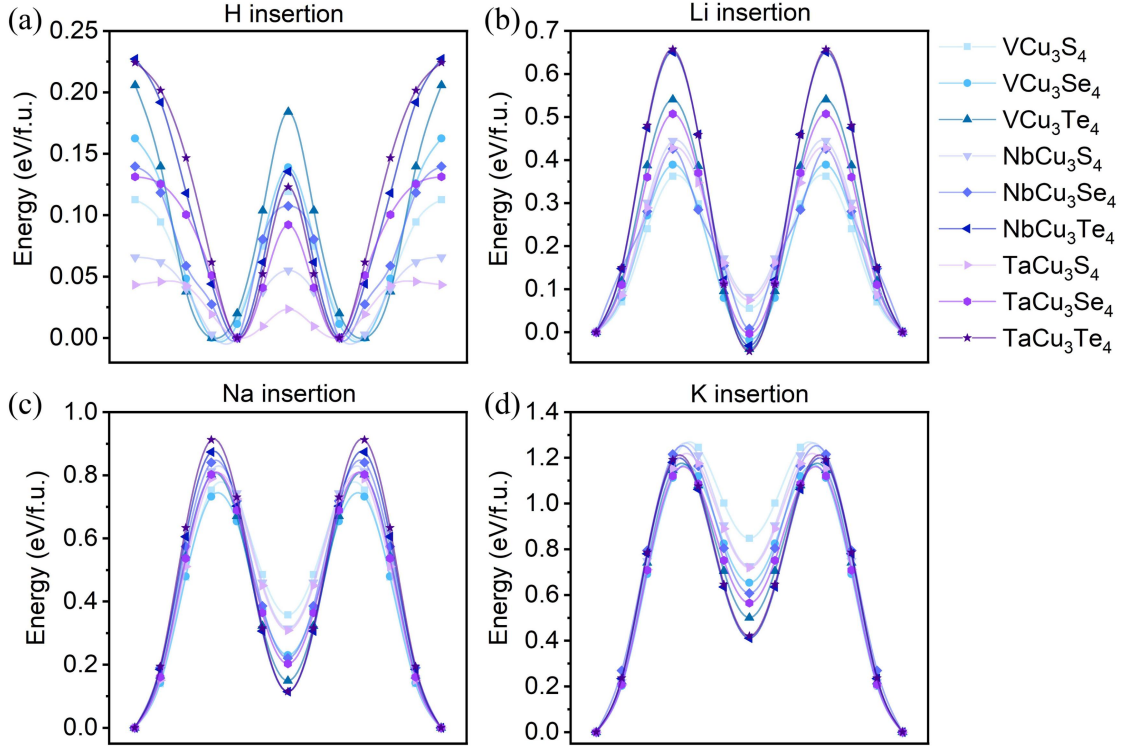

FIG. 3. Diffusion barriers of the inserted atoms (H/Li/Na/K) in $M$Cu$_3$X$_4$.

To evaluate the effect of inserted atom on the structural stability of $M$Cu$_3$X$_4$, we first performed a transition state search using the climbing image nudged elastic band method [47] to determine whether the central site is the lowest energy site of the inserted atom. The simulated diffusion process of the inserted atom in $M$Cu$_3$X$_4$ is shown in SM [46], and the calculated diffusion barrier is plotted in Fig. 3. The following two rules are found: (i) For the same material, the diffusion barrier aggrandizes gradually with the increase of the mass and radius of the inserted atom (H< Li< Na< K). (ii) For the same inserted atom, the increase of atomic mass and radius of $X$ atom (S< Se< Te) will also augment the diffusion barrier. Among them, the second rule is somewhat unexpected, because normally, the lattice constant of $M$Cu$_3$$X$$_4$ increases with the radius of $X$ atom, thus increasing the channel size and making the inserted atom easier to diffuse. We herein proposed an explanation. As shown in Fig. 4, we use $d_{X-X}$ to represent the distance between two $X$ atoms and find that $d_{X-X}$ does increase with the atomic number of $X$ (Table II), but meanwhile, the atomic radius of $X$ also increases. We set $l = d_{X-X} - 2r_X$ and measure the channel size by $l$. As shown in Table II, $l$ decreases with the increase of the atomic number of $X$, indicating that the increase of $r_X$ cancels out the increase of $d_{X-X}$, which adds the probability of the electron cloud overlapping between the inserted atom and $X$ atom, thus enhancing the interaction between them. As a result, the diffusion barrier of the inserted atom in $M$Cu$_3$$X$$_4$ increases with the atomic number of $X$.

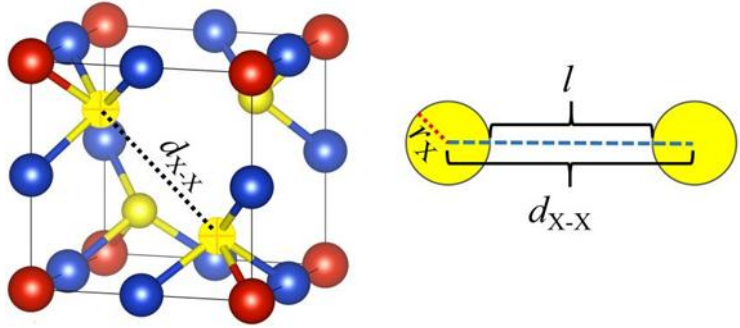

FIG. 4. Crystal structure of $M$Cu$_3$$X$$_4$ unit cell. $d_{X-X}$ is the distance between $X$ atom and $X$ atom, and $r_X$ is the radius of $X$ atom. The atomic radii of S, Se and Te are 1.09, 1.22 and 1.42 Å, respectively.

TABLE II. Calculated $d_{X-X}$ and $l$ in Fig. 4.

| Compound | VCu$_3$S$_4$ | VCu$_3$Se$_4$ | VCu$_3$Te$_4$ | NbCu$_3$S$_4$ | NbCu$_3$Se$_4$ | NbCu$_3$Te$_4$ | TaCu$_3$S$_4$ | TaCu$_3$Se$_4$ | TaCu$_3$Te$_4$ |
|---|---|---|---|---|---|---|---|---|---|
| $d_{X-X}$ (Å) | 4.067 | 4.123 | 4.204 | 4.048 | 4.063 | 4.147 | 4.063 | 4.065 | 4.160 |
| $l$ (Å) | 1.887 | 1.683 | 1.364 | 1.868 | 1.623 | 1.307 | 1.883 | 1.625 | 1.320 |

We also found that the lowest energy site for inserting H atom is not the central site as shown in Fig. 2(c). The results of structure re-optimization show that H-$M$Cu$_3$$X$$_4$ will transform into monoclinic crystal when the inserted H atom is located at the lowest energy site, see Fig. S3 [46]. While for the inserted Li/Na/K atom, there is a transition state, which is the face-center position of $M$Cu$_3$$X$$_4$ unit cell. As the inserted atomic radius enlarges, the energy difference between the transition state and the ground state gradually increases, among for Li-$M$Cu$_3$Te$_4$, the energy of transition state is lower than that of the initial structure.

We subsequently calculated the phonon spectra of $M$Cu$_3$$X$$_4$ after filling atoms, see Fig. 5 and Fig. S4 [46]. For the H atom insertion, the structure in Fig. S3 [46] is used. The results show that when inserting H atom, there is still a deep imaginary frequency in the phonon spectra even for H atom locating at the lowest energy position, suggesting that H-$M$Cu$_3$$X$$_4$ have dynamic instability. While for the insertion of Li/Na/K atom, the structure is adopted as shown in Fig. 2(c), and the stability of which would decline as the atomic radius and mass of Li/Na/K atom rise. Since Li atom has the proper atomic radius and mass, the phonon spectra of all Li insertion structures have no imaginary frequency, demonstrating that Li-$M$Cu$_3$X$_4$ have favorable dynamic stability, while K-$M$Cu$_3$X$_4$ have the opposite. Thus, we conclude that for H/Li/Na/K insertion $M$Cu$_3$X$_4$, Li-$M$Cu$_3$S$_4$ maybe the best choice as it has relatively low diffusion barrier while ensuring structural stability.

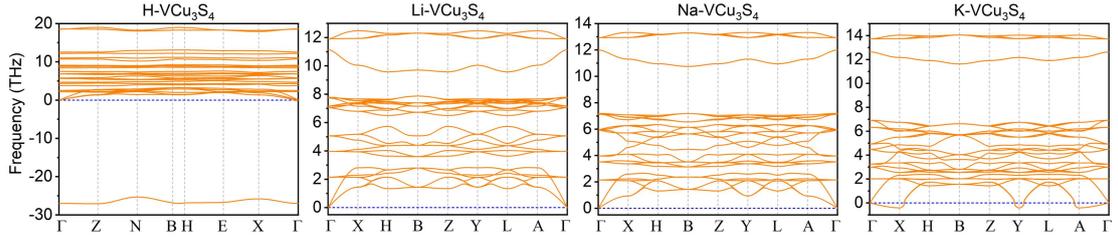

FIG. 5. Several representative phonon spectra of $M$Cu$_3$X$_4$ with inserting different atoms.

We have now confirmed that inserting Li atom into $M$Cu$_3$X$_4$ can still maintain excellent structural stability, which meets the requirements of nonvolatile and long cycle life of basic device units for neuromorphic computing. However, there is another requirement: two vastly different resistance states before and after inserting Li atom. Hence, the band structures of Li-$M$Cu$_3$X$_4$ are calculated as shown in Fig. 6(a) and Fig. S(5) [46], which shows that Li-$M$Cu$_3$X$_4$ all possess the band characteristics of metal with the Fermi level across the conduction band, meaning that the insertion of Li atom transforms $M$Cu$_3$X$_4$ from semiconductors to metals. Moreover, the insertion of Li atom has nearly no effect on the band dispersion of $M$Cu$_3$X$_4$, but only raises the position of Fermi level, which is similar to the band regulation achieved by electron doping [48]. We next verified the above inference by Bader charge analysis [49], which shows that 0.8~0.86 electrons per formula unit (f.u.) are transferred from Li to $M$Cu$_3$$X$$_4$, see Fig. S(6) [46], and it is this high concentration of electron doping that makes $M$Cu$_3$$X$$_4$ transform from semiconductors to metals. Recently, controllable hydrogenation accompanied by electron filling to achieve a tunable insulator-metal transition in VO$_2$ was reported [50], which provides strong support for our above analysis.

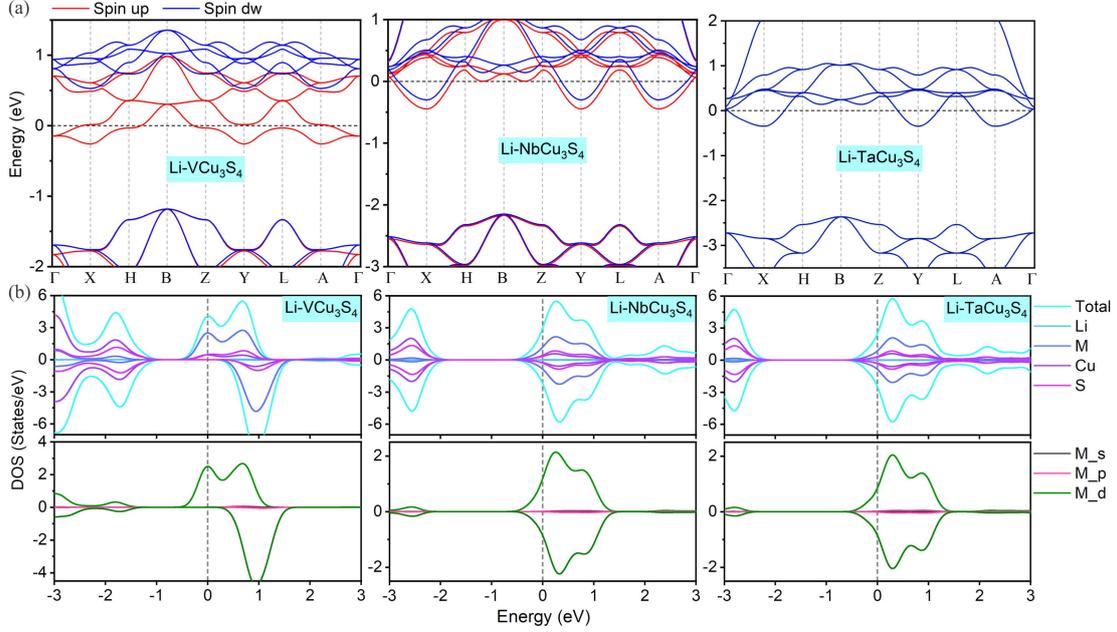

FIG. 6. (a) Band structures of Li-$M$Cu$_3$S$_4$. (b) Projected density of states of Li-$M$Cu$_3$S$_4$.

Additionally, we found that for the V, Nb and Ta elements, the V element compounds usually use PBE+U method to study the electronic structures [51]. Therefore, to confirm the influence of PBE+U for V atom on the results calculated in this work, we decided to take VCu$_3$S$_4$ as an example for testing. As shown in Fig. 7, after considering PBE+U for V atom, the band dispersion of $VCu_3S_4$ hardly changes, only the band gap increases, while Li-$VCu_3S_4$ still has the band characteristic of metal but shows the more obvious spin splitting. Thus, we know that although PBE+U for V atom will have an effect on the electronic structure of $VCu_3S_4$(Li-$VCu_3S_4$), it will not change the main conclusion of this paper, namely, the Li atom insertion will cause the transition of $VCu_3S_4$ from semiconductor to metal.

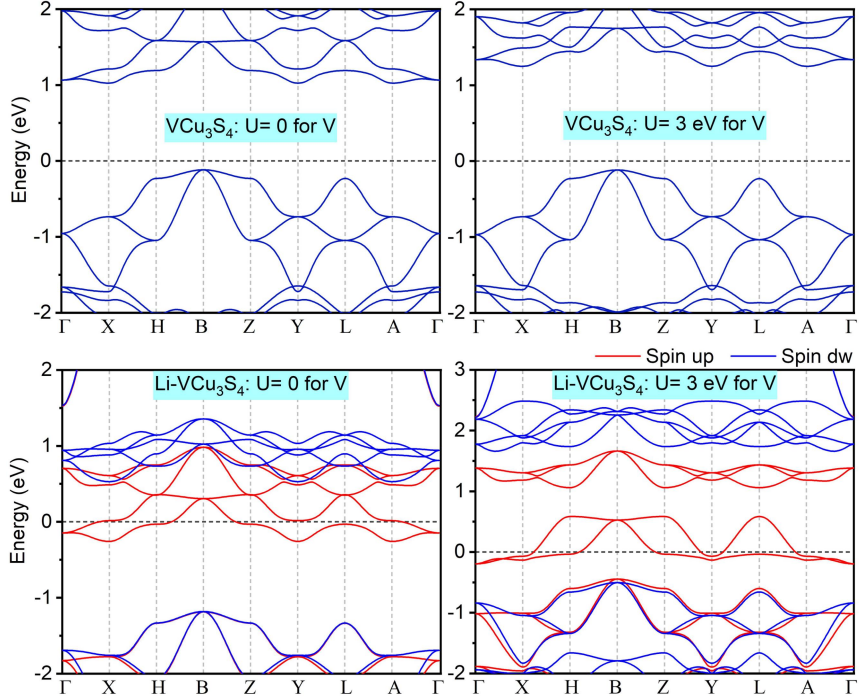

FIG. 7. Calculated band structures of $VCu_3S_4$ and Li-$VCu_3S_4$ with and without PBE+U method for V 3$d$ states.

Remarkably, Li-$VCu_3X_4$ and Li-$NbCu_3X_4$ appear spin polarization with a 1 $\mu_B$/f.u. and 0.25 $\mu_B$/f.u. average magnetic moment, respectively. Due to the obvious spin splitting (~0.8 eV) in the band of Li-$VCu_3X_4$, it has the band characteristics of half-metal [48,52], that is, only the spin-up bands cross the Fermi level. From the density of states in Fig. 6(b), the conduction band of Li-$M$Cu$_3$X$_4$ near the Fermi level is mainly contributed by $M$-$d$ states, and the degree of spin polarization of $M$-$d$ states gradually vanishes from left to right, corresponding to the change of magnetic moment, indicating that the magnetism originates from $M$-$d$ states. For this, we attribute it to the $M$ valence transition caused by electron filling accompanying with the insertion of Li atom.

For isolated $M$Cu$_3$X$_4$, $X$ is the sixth main family element, usually -2 valence in compounds; the valence electron of Cu is 3d$^{10}$4s$^1$, and that of $M$ is 3d$^3$4s$^2$, 4d$^4$5s$^1$ and 5d$^3$6s$^2$ for V, Nb and Ta, respectively. Therefore, we infer that the valence of $M$, Cu and $X$ for $M$Cu$_3$X$_4$ are respectively +5, +1 and -2 according to charge conservation, so there is no lone pair electron and consequently no magnetism in $M$Cu$_3$X$_4$. While Li atom is inserted into $M$Cu$_3$X$_4$, Li$^{+1}$[$M$Cu$_3$X$_4$]$^{-1}$ would be formed. From the charge transfer in Fig. S(6) [46], it can be seen that the electrons of Li are mainly transferred to $X$, while for $X^{2-}$, the outermost electron orbital is already full, and thus the electrons from Li cannot be stably bound and may be acquired by $M$. The probability of $M$ acquiring these electrons can be estimated according to the following ionicity formula defined by the Paling, $f_i = 1 - \exp[-(x_A - x_B)^2/4]$, where $x_A$ and $x_B$ represent the electronegativity of atom A and atom B, respectively. V/Nb/Ta is the fifth subgroup element, the electronegativity of which weakens with the increase of their atomic number, corresponding to the enhanced ionicity between *M* and *X*. The stronger ionicity, the harder it is to transfer electrons from *X* to *M*. Thus for V/Nb/Ta, V atom gains the most electrons from *X* and possibly changes from $V^{5+}$ to $V^{4+}$, resulting in a 1 $\mu_B$/f.u. magnetic moment; Nb atom gets fewer electrons from *X* and thus has a smaller magnetic moment; Ta atom recovers almost no electrons from *X*, so there is no spin polarization. In addition, we also compared the energy of ferromagnetic order and antiferromagnetic order of Li-VCu$_3$X$_4$ and confirmed that the ferromagnetic state is the energy ground state, see Fig. S7 and Table S2 [46].

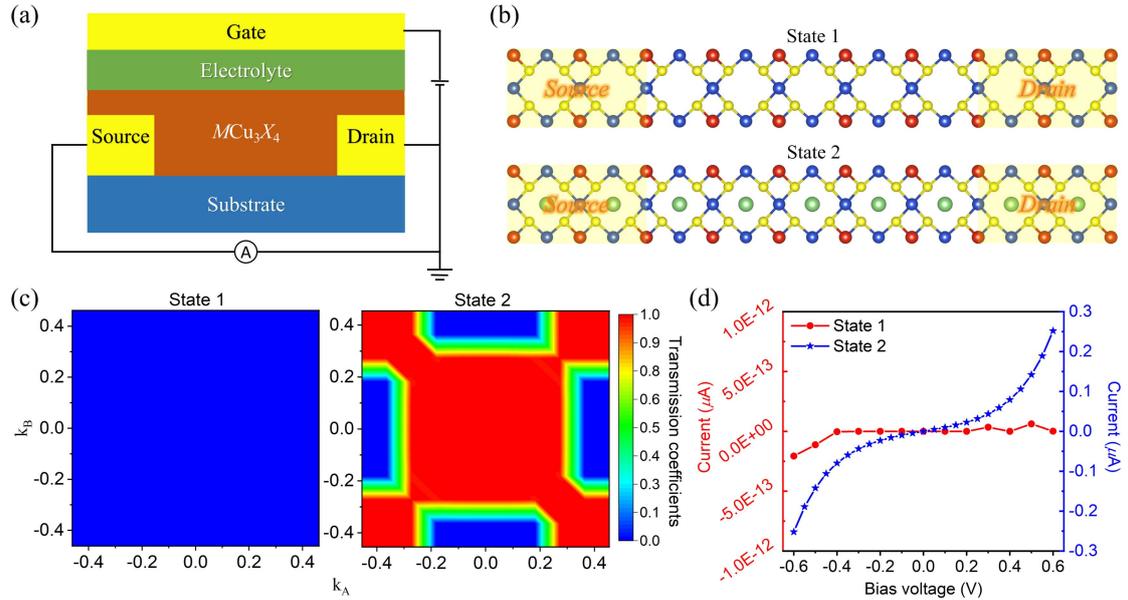

FIG. 8. (a) Schematic of the ionic synaptic transistor based on *M*Cu$_3$X$_4$. (b) Two ideal states of the device as shown in (a) are realized by adjusting the lithium atoms embedding and stripping through the gate voltage. (c) Calculated zero-bias transmission coefficients at the Fermi level of device states 1 and 2 as shown in (b). (d) Calculated *I-V* curves of device states 1 and 2 as shown in (b).

Through the above research, we basically confirm that *M*Cu$_3$X$_4$ can be used to design ion synaptic transistors. The device model is shown in Fig. 8(a), and its working principle is the same as that in Ref. 11. To further evaluate the device performance, we take $VCu_3S_4$ as an example and quantitatively analyze the electrical transport properties of device. Fig. 8(b) shows two ideal states of the device, which can be achieved by controlling Li insertion/extraction. The transmission coefficient can represent the probability of a particle tunneling through a barrier. Fig. 8(c) plots the zero-bias transmission coefficients of state 1 and state 2 at the Fermi level, which are close to 0 and 1 for state 1 and state 2, respectively, corresponding to the semiconducting feature of $VCu_3S_4$ and metallic feature of $Li-VCu_3S_4$. We next plot the volt-ampere characteristic curves of state 1 and state 2 in Fig. 8(d). The results show that under the same bias range (-0.6 V~0.6 V), the current of state 1 is in the order of $10^{-13}$ $\mu A$, while that of state 2 is in the order of 0.1 $\mu A$, indicating that the resistance ratio of two states reaches a surprising order of $10^{12}$. Finally, we also consider the electron structures of the partial filling states for inserting Li into $VCu_3S_4$. The results show that $VCu_3S_4$ can achieve the semiconductor-metal transition at a lower inserted Li concentration (0.46%~1.54%), see SM [46].

## IV. SUMMARY AND DISSCUSSION

In summary, we proposed a promising class of candidate materials $MCu_3X_4$ for neuromorphic computing through first-principles simulations.

When Li atom is inserted into $M$Cu$_3$$X$$_4$ to form Li-$M$Cu$_3$$X$$_4$, it not only maintains good structural stability, but also transforms from semiconductors to metals. Meanwhile, the low diffusion barrier of Li atom in $M$Cu$_3$$X$$_4$ ensures the feasibility of gated Li insertion/extraction. These results show that $M$Cu$_3$$X$$_4$ can be used to design ionic synaptic transistors and provide inspiration for the search of candidate materials related to neuromorphic computing by theoretical calculations. It is believed that with the upgrading of computers and algorithms, self-optimization and upgrading of artificial intelligence with autonomous learning ability based on neuromorphic computing will become a reality in the future. Last but not least, there are still some details of this work that deserve further investigation. For example, Li-VCu$_3$$X$$_4$ with half-metallicity may have potential applications in spintronic devices and whether Li-$M$Cu$_3$$X$$_4$ may be used in solid-state ion batteries [53].

## ACKNOWLEDGEMENTS

This work was supported by National Key R&D Program of China (No. 2018YFA0703700), the National Natural Science Foundation of China (Nos. 91964203, 62274121, 62104171, 62104172, and 62004142), the Strategic Priority Research Program of Chinese Academy of Sciences (No. XDB44000000), the Natural Science Foundation of Hubei Province, China (No. 2021CFB037), and the Fundamental Research Funds for the Central Universities (No. 2042021kf0067). The numerical calculations in this paper have been done on the supercomputing system in the Supercomputing Center of Wuhan University.

[1] H. Lasi, P. Fettke, H.-G. Kemper, T. Feld, and M. Hoffmann, Industry 4.0, Bus. Inf. Syst. Eng. **6**, 239 (2014).

[2] S. H. Jo, T. Chang, I. Ebong, B. B. Bhadviya, P. Mazumder, and W. Lu, Nanoscale Memristor Device as Synapse in Neuromorphic Systems, Nano Lett. **10**, 1297 (2010).

[3] J. J. Yang, D. B. Strukov, and D. R. Stewart, Memristive Devices for Computing, Nat. Nanotechnol. **8**, 13 (2013).

[4] Z. Wang, S. Joshi, S. E. Savel'ev, H. Jiang, R. Midya, P. Lin, M. Hu, N. Ge, J. P. Strachan, Z. Li, *et al.*, Memristors with Diffusive Dynamics as Synaptic Emulators for Neuromorphic Computing, Nat. Mater. **16**, 101 (2017).

[5] Y. van de Burgt, E. Lubberman, E. J. Fuller, S. T. Keene, G. C. Faria, S. Agarwal, M. J. Marinella, A. Alec Talin, and A. Salleo, A Non-Volatile Organic Electrochemical Device as a Low-Voltage Artificial Synapse for Neuromorphic Computing, Nat. Mater. **16**, 414 (2017).

[6] C. Cortes and V. Vapnik, Support-Vector Networks, Mach. Learn. **20**, 273 (1995).

[7] F. Pedregosa, G. Varoquaux, A. Gramfort, V. Michel, B. Thirion, O.

Grisel, M. Blondel, P. Prettenhofer, R. Weiss, V. Dubourg, *et al.*, Scikit-Learn: Machine Learning in Python, J. Mach. Learn. Res. **12**, 2825 (2011).

[8] Y. LeCun, Y. Bengio, and G. Hinton, Deep Learning, Nature **521**, 436 (2015).

[9] V. Mnih, K. Kavukcuoglu, D. Silver, A. A. Rusu, J. Veness, M. G. Bellemare, A. Graves, M. Riedmiller, A. K. Fidjeland, G. Ostrovski, *et al.*, Human-Level Control through Deep Reinforcement Learning, Nature **518**, 529 (2015).

[10] X. Zhu, D. Li, X. Liang, and W. D. Lu, Ionic Modulation and Ionic Coupling Effects in $MoS_2$ Devices for Neuromorphic Computing, Nat. Mater. **18**, 141 (2019).

[11] E. J. Fuller, F. E. Gabaly, F. Léonard, S. Agarwal, S. J. Plimpton, R. B. Jacobs-Gedrim, C. D. James, M. J. Marinella, and A. A. Talin, Li-Ion Synaptic Transistor for Low Power Analog Computing, Adv. Mater. **29**, 1604310 (2017).

[12] J. Zhu, Y. Yang, R. Jia, Z. Liang, W. Zhu, Z. U. Rehman, L. Bao, X. Zhang, Y. Cai, L. Song, *et al.*, Ion Gated Synaptic Transistors Based on 2D van Der Waals Crystals with Tunable Diffusive Dynamics, Adv. Mater. **30**, 1800195 (2018).

[13] D. Kireev, S. Liu, H. Jin, T. Patrick Xiao, C. H. Bennett, D. Akinwande, and J. A. C. Incorvia, Metaplastic and Energy-Efficient

Biocompatible Graphene Artificial Synaptic Transistors for Enhanced Accuracy Neuromorphic Computing, Nat. Commun. **13**, 4386 (2022).

[14] B. Zhao and J. Ravichandran, Low-Power Microwave Relaxation Oscillators Based on Phase-Change Oxides for Neuromorphic Computing, Phys. Rev. Appl. **11**, 014020 (2019).

[15] Y. Shi and L.-Q. Chen, Intrinsic Insulator-Metal Phase Oscillations, Phys. Rev. Appl. **17**, 014042 (2022).

[16] C. Adda, M.-H. Lee, Y. Kalcheim, P. Salev, R. Rocco, N. M. Vargas, N. Ghazikhanian, C.-P. Li, G. Albright, M. Rozenberg, *et al.*, Direct Observation of the Electrically Triggered Insulator-Metal Transition in $V_3O_5$ Far below the Transition Temperature, Phys. Rev. X **12**, 011025 (2022).

[17] G. Li, D. Xie, H. Zhong, Z. Zhang, X. Fu, Q. Zhou, Q. Li, H. Ni, J. Wang, E. Guo, *et al.*, Photo-Induced Non-Volatile $VO_2$ Phase Transition for Neuromorphic Ultraviolet Sensors, Nat. Commun. **13**, 1729 (2022).

[18] M. S. Nikoo, R. Soleimanzadeh, A. Krammer, G. M. Marega, Y. Park, J. Son, A. Schueler, A. Kis, P. J. W. Moll, and E. Matioli, Electrical Control of Glass-like Dynamics in Vanadium Dioxide for Data Storage and Processing, Nat. Electron. **5**, 596 (2022).

[19] L. Xu, H. Xiong, Z. Fu, M. Deng, S. Wang, J. Zhang, L. Shang, K.

Jiang, Y. Li, L. Zhu, *et al.*, High Conductance Margin for Efficient Neuromorphic Computing Enabled by Stacking Nonvolatile van der Waals Transistors, Phys. Rev. Appl. **16**, 044049 (2021).

[20] Y. Wang, W. Li, Y. Guo, X. Huang, Z. Luo, S. Wu, H. Wang, J. Chen, X. Li, X. Zhan, *et al.*, A Gate-Tunable Artificial Synapse Based on Vertically Assembled van der Waals Ferroelectric Heterojunction, J. Mater. Sci. Technol. **128**, 239 (2022).

[21] D. Prychynenko, M. Sitte, K. Litzius, B. Krüger, G. Bourianoff, M. Kläui, J. Sinova, and K. Everschor-Sitte, Magnetic Skyrmion as a Nonlinear Resistive Element: A Potential Building Block for Reservoir Computing, Phys. Rev. Appl. **9**, 014034 (2018).

[22] R. Chen, C. Li, Y. Li, J. J. Miles, G. Indiveri, S. Furber, V. F. Pavlidis, and C. Moutafis, Nanoscale Room-Temperature Multilayer Skyrmionic Synapse for Deep Spiking Neural Networks, Phys. Rev. Appl. **14**, 014096 (2020).

[23] S. Zhang and Y. Tserkovnyak, Antiferromagnet-Based Neuromorphics Using Dynamics of Topological Charges, Phys. Rev. Lett. **125**, 207202 (2020).

[24] D. Marković, M. W. Daniels, P. Sethi, A. D. Kent, M. D. Stiles, and J. Grollier, Easy-Plane Spin Hall Nano-Oscillators as Spiking Neurons for Neuromorphic Computing, Phys. Rev. B **105**, 014411 (2022).

[25] M. G. Vergniory, L. Elcoro, C. Felser, N. Regnault, B. A. Bernevig,

and Z. Wang, A Complete Catalogue of High-Quality Topological Materials, Nature **566**, 480 (2019).

[26] Y. Xu, L. Elcoro, Z.-D. Song, B. J. Wieder, M. G. Vergniory, N. Regnault, Y. Chen, C. Felser, and B. A. Bernevig, High-Throughput Calculations of Magnetic Topological Materials, Nature **586**, 702 (2020).

[27] N. Regnault, Y. Xu, M.-R. Li, D.-S. Ma, M. Jovanovic, A. Yazdani, S. S. P. Parkin, C. Felser, L. M. Schoop, N. Phuan Ong, *et al.*, Catalogue of Flat-Band Stoichiometric Materials, Nature **603**, 824 (2022).

[28] A. B. Kehoe, The Electronic Structure of Sulvanite Structured Semiconductors $Cu_3MCh_4$ (M = V, Nb, Ta; Ch = S, Se, Te): Prospects for Optoelectronic Applications, J. Mater. Chem. C **3**, 12236 (2015).

[29] Y. Li, M. Wu, T. Zhang, X. Qi, G. Ming, G. Wang, X. Quan, and D. Yang, Natural Sulvanite $Cu_3MX_4$ (M = Nb, Ta; X = S, Se): Promising Visible-Light Photocatalysts for Water Splitting, Comput. Mater. Sci. **165**, 137 (2019).

[30] V. Mantella, S. Ninova, S. Saris, A. Loiudice, U. Aschauer, and R. Buonsanti, Synthesis and Size-Dependent Optical Properties of Intermediate Band Gap $Cu_3VS_4$ Nanocrystals, Chem. Mater. **31**, 532 (2019).

[31] M. Liu, C.-Y. Lai, G. S. Selopal, and D. R. Radu, Synthesis and

Optoelectronic Properties of Cu$_3$VSe$_4$ Nanocrystals, PLOS ONE **15**, e0232184 (2020).

[32] E. Haque, Outstanding Thermoelectric Performance of MCu$_3$X$_4$ (M = V, Nb, Ta; X = S, Se, Te) with Unaffected Band Degeneracy under Pressure, ACS Appl. Energy Mater. **4**, 1942 (2021).

[33] J. Wen, H. Huang, X. Yu, D. Wang, K. Guo, D. Wan, J. Luo, and J.-T. Zhao, Thermoelectric Properties of P-Type Cu$_3$VSe$_4$ with High Seebeck Coefficients, J. Alloys Compd. **879**, 160387 (2021).

[34] H. W. Kroto, The Stability of the Fullerenes C$_n$, with n = 24, 28, 32, 36, 50, 60 and 70, Nature **329**, 529 (1987).

[35] A. A. Popov, S. Yang, and L. Dunsch, Endohedral Fullerenes, Chem. Rev. **113**, 5989 (2013).

[36] N. A. Romero, J. Kim, and R. M. Martin, Electronic Structures and Superconductivity of Endohedrally Doped C$_{28}$ Solids from First Principles, Phys. Rev. B **76**, 205405 (2007).

[37] J. Li and R. Wu, Two-Dimensional Multifunctional Materials from Endohedral Fullerenes, Phys. Rev. B **103**, 115417 (2021).

[38] H. Arribart, B. Sapoval, O. Gorochov, and N. LeNagard, Fast Ion Transport at Room Temperature in the Mixed Conductor Cu$_3$VS$_4$, Solid State Commun. **26**, 435 (1978).

[39] G. Kresse and J. Furthmüller, Efficiency of Ab-Initio Total Energy Calculations for Metals and Semiconductors Using a Plane-Wave

Basis Set, Comput. Mater. Sci. **6**, 15 (1996).

[40] J. P. Perdew, K. Burke, and M. Ernzerhof, Generalized Gradient Approximation Made Simple, Phys. Rev. Lett. **77**, 3865 (1996).

[41] S. L. Dudarev, G. A. Botton, S. Y. Savrasov, C. J. Humphreys, and A. P. Sutton, Electron-Energy-Loss Spectra and the Structural Stability of Nickel Oxide: An LSDA+U Study, Phys. Rev. B **57**, 1505 (1998).

[42] P. E. Blöchl, Projector Augmented-Wave Method, Phys. Rev. B **50**, 17953 (1994).

[43] H. J. Monkhorst and J. D. Pack, Special Points for Brillouin-Zone Integrations, Phys. Rev. B **13**, 5188 (1976).

[44] M. Brandbyge, J.-L. Mozos, P. Ordejón, J. Taylor, and K. Stokbro, Density-Functional Method for Nonequilibrium Electron Transport, Phys. Rev. B **65**, 165401 (2002).

[45] J. Taylor, H. Guo, and J. Wang, Ab Initio Modeling of Quantum Transport Properties of Molecular Electronic Devices, Phys. Rev. B **63**, 245407 (2001).

[46] See the Supplemental Material at xxx for the band structures of isolated $M$Cu$_3$$X$$_4$ and Li-$M$Cu$_3$$X$$_4$, nudged elastic band simulations of H/Li/Na/K in $M$Cu$_3$$X$$_4$, structural parameters of H-$M$Cu$_3$$X$$_4$, phonon spectra of $M$Cu$_3$$X$$_4$ with inserting different atoms, charge transfer in Li-$M$Cu$_3$$X$$_4$, ferromagnetic order and antiferromagnetic order of Li-VCu$_3$$X$$_4$, and the electron structures of the partial filling states for

inserting Li into VCu$_3$S$_4$.

[47] G. Henkelman, B. P. Uberuaga, and H. Jónsson, A Climbing Image Nudged Elastic Band Method for Finding Saddle Points and Minimum Energy Paths, J. Chem. Phys. **113**, 9901 (2000).

[48] H. Wang, F. Fan, S. Zhu, and H. Wu, Doping Enhanced Ferromagnetism and Induced Half-Metallicity in CrI$_3$ Monolayer, EPL Europhys. Lett. **114**, 47001 (2016).

[49] G. Henkelman, A. Arnaldsson, and H. Jónsson, A Fast and Robust Algorithm for Bader Decomposition of Charge Density, Comput. Mater. Sci. **36**, 354 (2006).

[50] L. Li, M. Wang, Y. Zhou, Y. Zhang, F. Zhang, Y. Wu, Y. Wang, Y. Lyu, N. Lu, G. Wang, *et al.*, Manipulating the Insulator-Metal Transition through Tip-Induced Hydrogenation, Nat. Mater. **21**, 1246 (2022).

[51] X. Liu, A. P. Pyatakov, and W. Ren, Magnetoelectric Coupling in Multiferroic Bilayer VS$_2$, Phys. Rev. Lett. **125**, 247601 (2020).

[52] Y. Zhao, J. Zhang, S. Yuan, and Z. Chen, Nonvolatile Electrical Control and Heterointerface-Induced Half-Metallicity of 2D Ferromagnets, Adv. Funct. Mater. **29**, 1901420 (2019).

[53] W. Wang, Y. Gang, Z. Hu, Z. Yan, W. Li, Y. Li, Q.-F. Gu, Z. Wang, S.-L. Chou, H.-K. Liu, *et al.*, Reversible Structural Evolution of Sodium-Rich Rhombohedral Prussian Blue for Sodium-Ion Batteries,

Nat. Commun. **11**, 980 (2020).

# Supplementary Information

# First-principles Prediction of Potential Candidate Materials $M$Cu$_3$X$_4$ ($M$ = V, Nb, Ta; $X$ = S, Se, Te) for Neuromorphic Computing

Baoxing Zhai,[1] Ruiqing Cheng,[1] Tianxing Wang,[2] Li Liu,[1] Lei Yin,[1] Yao Wen,[1] Hao Wang,[1] Sheng Chang,[1] and Jun He[1,3,*]

[1]*Key Laboratory of Artificial Micro- and Nano-structures of Ministry of Education, and School of Physics and Technology, Wuhan University, Wuhan 430072, China*
[2]*School of physics, Henan Normal University, Xinxiang 453007, China*
[3]*Wuhan Institute of Quantum Technology, Wuhan 430206, China*

[*]he-jun@whu.edu.cn

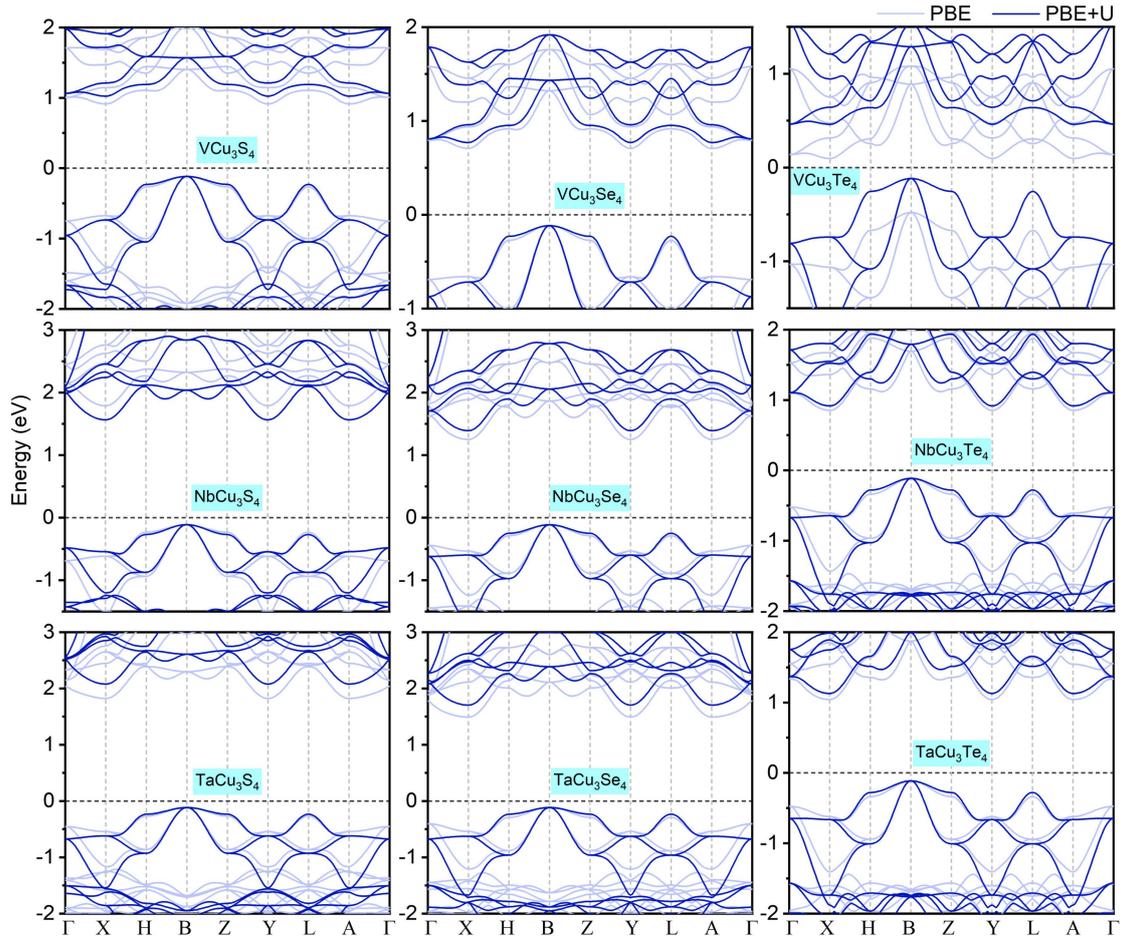

FIG. S1. Band structures of $M\mathrm{Cu}_3X_4$ ($M$ = V, Nb, Ta; $X$ = S, Se, Te).

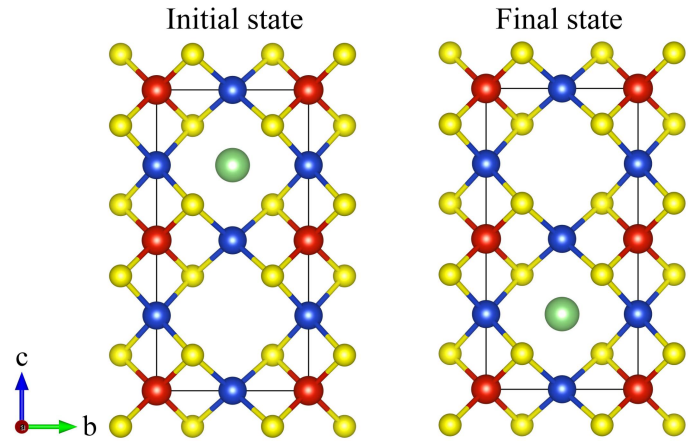

FIG. S2. Initial and final state structures of the transition state search. First, the structure after inserting atoms is expanded into a 1×1×2 supercell, and then the upper inserted atom and the lower inserted atom is removed respectively to obtain the initial and final state structure.

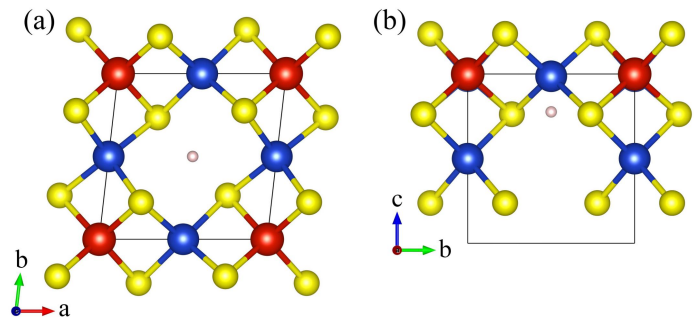

FIG. S3. (a) Top view and (b) side view of the lowest energy crystal structure after hydrogen atom insertion into $M\text{Cu}_3X_4$. The detailed lattice parameters of H-$M\text{Cu}_3X_4$ are shown in Table S1.

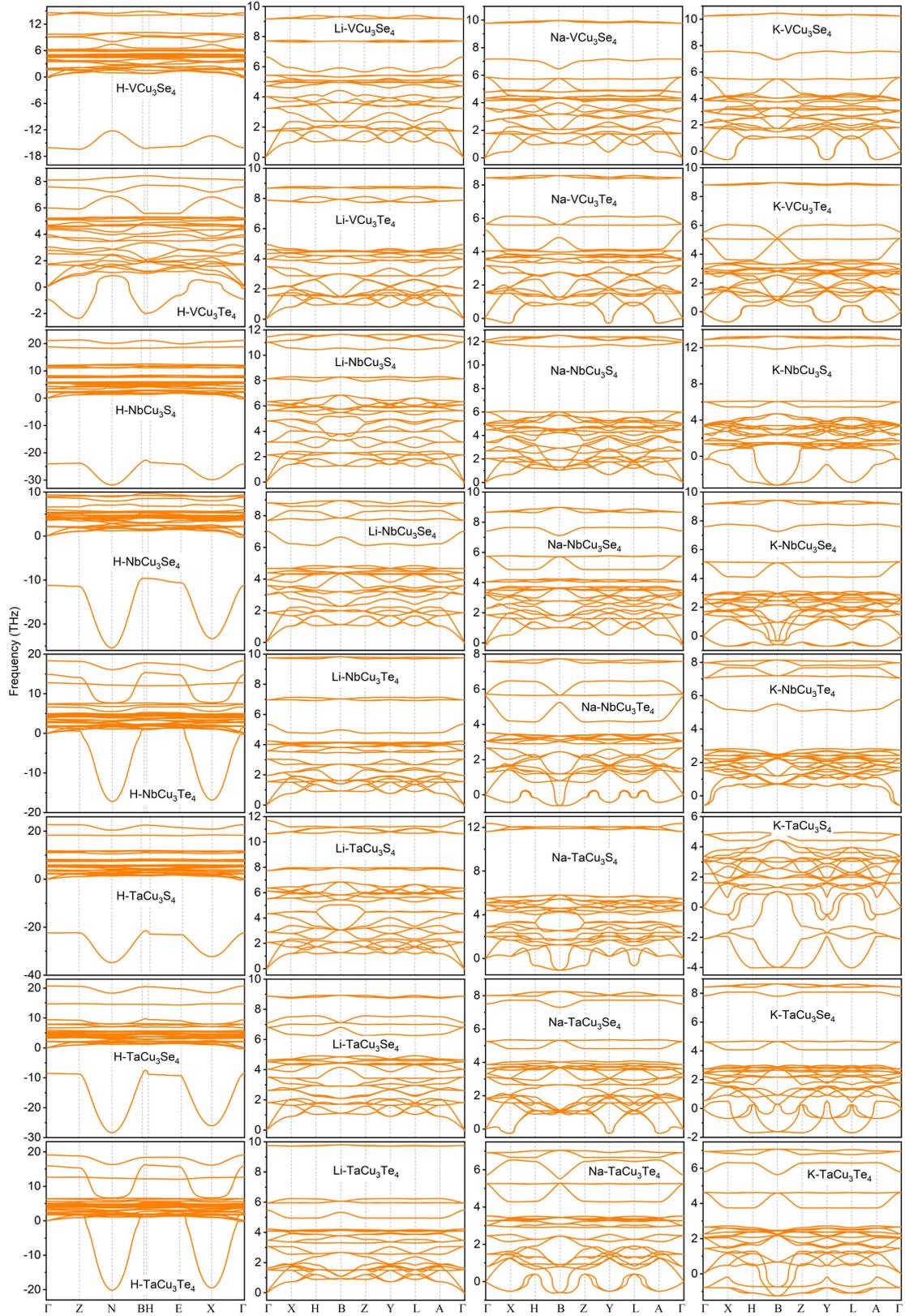

FIG. S4. Phonon spectra of $M$Cu$_3$$X$$_4$ with inserting different atoms as a supplement to Fig. 5.

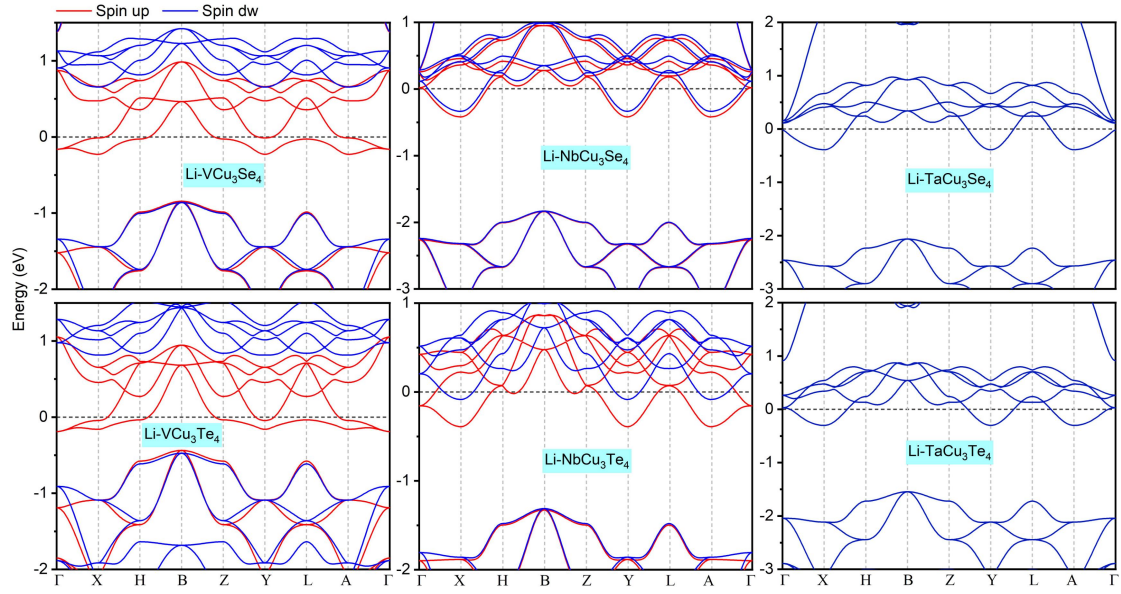

FIG. S5. Band structures of Li-$M$Cu$_3$X$_4$ as a supplement to Fig. 6(a).

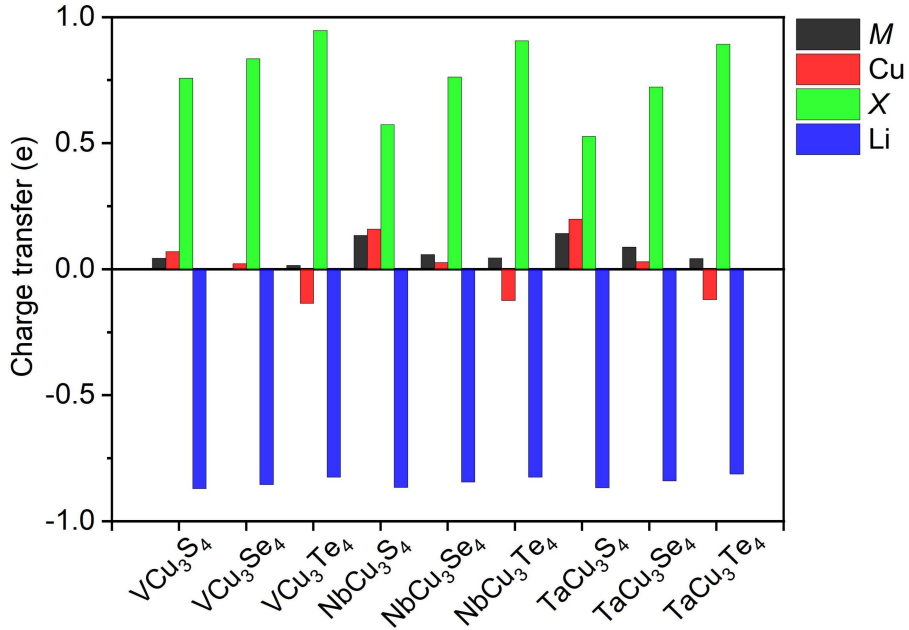

FIG. S6. Charge transfer distribution of Li atom inserted into $M$Cu$_3$X$_4$.

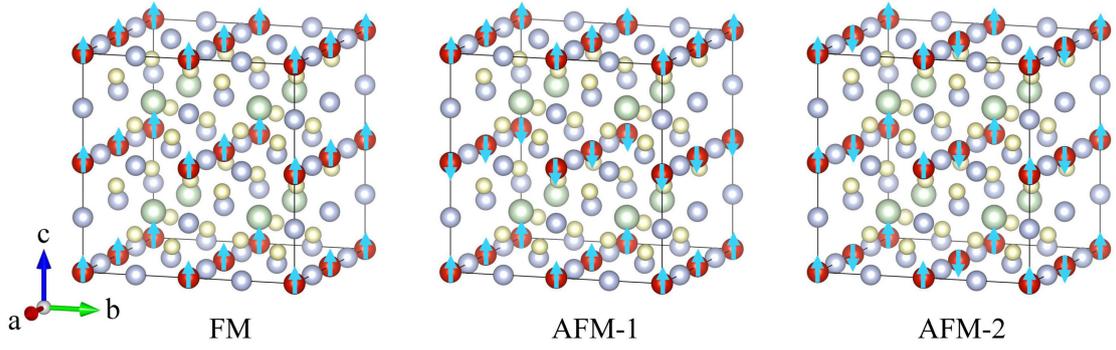

FIG. S7. Ferromagnetic (FM) order and antiferromagnetic (AFM) order of Li-VCu$_3$X$_4$.

TABLE S1. Re-optimized lattice parameters of H-$M$Cu$_3$X$_4$.

| Compound | $a = b$ (Å) | $c$ (Å) | $\alpha = \beta$ (°) | $\gamma$ (°) |
|---|---|---|---|---|
| H-VCu$_3$S$_4$ | 5.398 | 5.480 | 90 | 83.519 |
| H-VCu$_3$Se$_4$ | 5.613 | 5.710 | 90 | 85.819 |
| H-VCu$_3$Te$_4$ | 5.907 | 6.070 | 90 | 88.284 |
| H-NbCu$_3$S$_4$ | 5.527 | 5.502 | 90 | 84.290 |
| H-NbCu$_3$Se$_4$ | 5.730 | 5.698 | 90 | 86.111 |
| H-NbCu$_3$Te$_4$ | 6.012 | 5.970 | 90 | 88.282 |
| H-TaCu$_3$S$_4$ | 5.539 | 5.511 | 90 | 80.085 |
| H-TaCu$_3$Se$_4$ | 5.729 | 5.693 | 90 | 86.150 |
| H-TaCu$_3$Te$_4$ | 6.014 | 5.966 | 90 | 88.365 |

TABLE S2. Energies of FM order and AFM order of Li-VCu$_3$X$_4$.

| Compound | $E_{FM}$ (eV/f.u.) | $E_{AFM1}$ (eV/f.u.) | $E_{AFM2}$ (eV/f.u.) |
|---|---|---|---|
| Li-VCu$_3$S$_4$ | -44.465 | -44.454 | -44.454 |
| Li-VCu$_3$Se$_4$ | -40.849 | -40.832 | -40.832 |
| Li-VCu$_3$Te$_4$ | -37.245 | -37.224 | -37.224 |

**Electronic structures of the partial filling states for inserting Li into $VCu_3S_4$:**

By expanding $VCu_3S_4$ into 2×2×2 and 3×3×3 supercells and setting the number of Li atom in them, we obtain different partial filling states, see Fig. S8, and the corresponding band structures are displayed in Fig. S9 and Fig. S10. We find that as the concentration of inserted Li atom increases, the Fermi level is gradually raised from being in the forbidden band to crossing the conduction band, which represents the transition from semiconductor to metal. By calculating the Li concentration corresponding to $Li(VCu_3S_4)_{27}$ and $Li(VCu_3S_4)_8$, we realize that the concentration of inserted Li that causes the semiconductor-metal transition of $VCu_3S_4$ is 0.46%~1.54%.

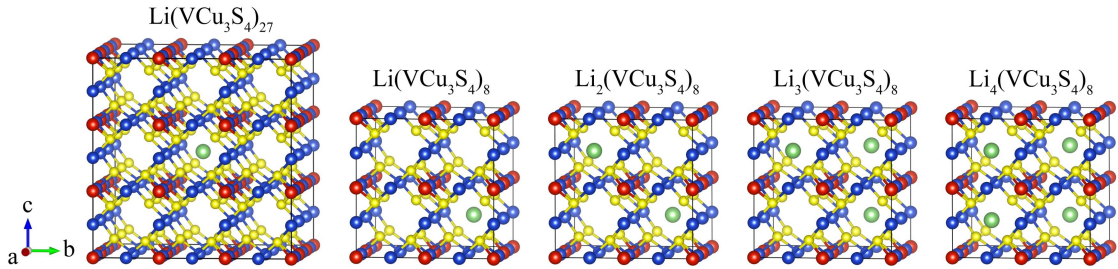

FIG. S8. Crystal structures of $Li(VCu_3S_4)_{27}$, $Li(VCu_3S_4)_8$, $Li_2(VCu_3S_4)_8$, $Li_3(VCu_3S_4)_8$ and $Li_4(VCu_3S_4)_8$.

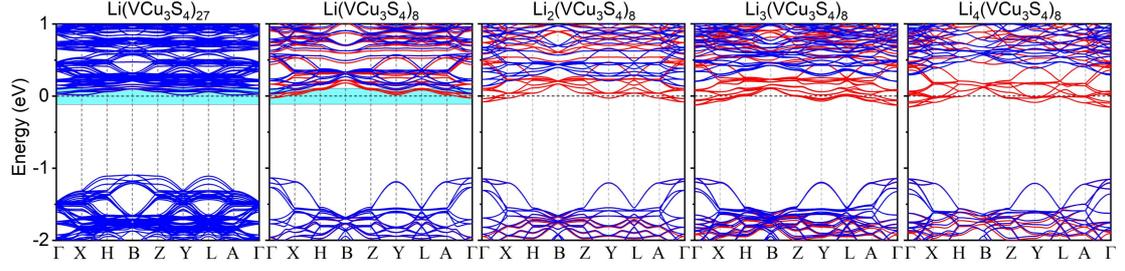

FIG. S9. Calculated band structures of Li(VCu$_3$S$_4$)$_{27}$, Li(VCu$_3$S$_4$)$_8$, Li$_2$(VCu$_3$S$_4$)$_8$, Li$_3$(VCu$_3$S$_4$)$_8$ and Li$_4$(VCu$_3$S$_4$)$_8$. The red and blue lines denote the spin-up and spin-down channels, respectively. The Fermi level is set as zero.

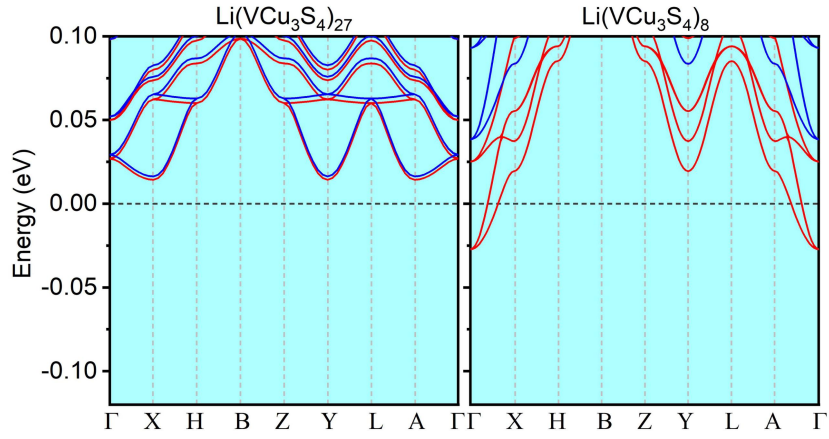

FIG. S10. Enlargement of cyan region in band structures of Li(VCu$_3$S$_4$)$_{27}$ and Li(VCu$_3$S$_4$)$_8$ in Fig. S9.